\documentclass[journal]{IEEEtran}

\usepackage[utf8]{inputenc} 
\usepackage[T1]{fontenc}
\usepackage{color}
\usepackage{float}
\usepackage{amsbsy}
\usepackage{amstext}
\usepackage{setspace}
\usepackage{orcidlink}
\usepackage{graphicx}
\usepackage{bm}
\usepackage{float}
\usepackage{amsmath, amssymb}
\usepackage{gensymb}
\usepackage{upgreek}
\usepackage{breqn}
\usepackage{blindtext}
\usepackage{packages/widetext}
\usepackage{packages/dblfloatfix}
\usepackage{subcaption}
\usepackage{tabularx}  

\begin{document}

\title{
Engineering the Frozen Mode Near a Stationary Inflection Point in Coupled Single-Mode Waveguides with Lateral Gratings}

\author{Albert~Herrero-Parareda~\orcidlink{https://orcid.org/0000-0002-8501-5775}, 
        Nathaniel~Furman~\orcidlink{https://orcid.org/0000-0001-7896-2929},
        Bradley~J.~Thompson\orcidlink{https://orcid.org/0000-0002-7416-2112},
        Ricky~Gibson\orcidlink{https://orcid.org/0000-0002-2567-6707},
        Ilya~Vitebskiy\orcidlink{https://orcid.org/0000-0001-8375-2088},
        and~Filippo~Capolino~\orcidlink{https://orcid.org/0000-0003-0758-6182}
\thanks{A.~Herrero-Parareda, N.~Furman, and F.~Capolino are affiliated with the Department of Electrical Engineering and Computer Science, University of California, Irvine,
CA, 92697 USA. B.~J.~Thompson, R.~Gibson, and I.~Vitebskiy are affiliated with Air Force Research Laboratory, Sensors Directorate, Wright-Patterson AFB, OH, 45433 USA. e-mail: f.capolino@uci.edu.}
}

\maketitle


\begin{abstract}
We introduce a systematic approach to designing slow-light photonic integrated circuits supporting a frozen mode associated with a third-order exceptional point of degeneracy (EPD), known as a stationary inflection point (SIP). The proposed approach employs three-way coupled waveguides incorporating lateral gratings and operating at telecommunication wavelengths. We investigate two distinct waveguide configurations and demonstrate the formation of SIPs through full-wave eigenmode simulations. We also analyze the sensitivity of the dispersion characteristics to geometric perturbations and realistic fabrication constraints. A periodic waveguide with integrated tapered terminations is fabricated and experimentally characterized, revealing spectral features that are qualitatively consistent with numerical predictions. Simulations of finite-length structures further demonstrate substantial group-delay enhancement near the SIP frequency. These results establish a practical approach to realizing frozen-mode waveguides and integrating SIP-based slow-light functionalities into silicon photonic circuits.
\end{abstract}

\section{Introduction}
\label{ch:intro}
Advances in nanoscale fabrication technologies have enabled the development of increasingly sophisticated photonic integrated circuits (PICs), in which optical components such as waveguides and resonators are integrated on a common semiconductor substrate.

In this work, we investigate a new configuration of passive PIC waveguides consisting of three coupled single-mode waveguides incorporating lateral gratings, designed to support an exceptional point of degeneracy (EPD). An EPD is a point in the parameter space of a system at which two or more eigenstates coalesce. Early studies of guiding structures supporting this class of degeneracies were presented in Refs.~\cite{figotin_oblique_2003,figotin_electromagnetic_2003, figotin_frozen_2006, figotin_slow_2011}, where the underlying physical phenomena were investigated in terms of stationary points of the dispersion relation, without explicitly invoking the concept of exceptional points. Here, we focus on the stationary inflection point (SIP), a third-order EPD associated with the coalescence of three eigenmodes. EPDs arise in a broad range of classical and quantum systems \cite{heiss_exceptional_2004} and have attracted considerable attention in systems with balanced gain and loss exhibiting parity-time (PT) symmetry \cite{ruter_observation_2010, ramezani_unidirectional_2010, othman_theory_2017}.

EPDs in periodic photonic structures can be realized through dispersion engineering without requiring gain or loss \cite{nada_theory_2017}. The underlying physical concepts were investigated in earlier studies \cite{figotin_gigantic_2005, figotin_slow_2006, figotin_slow_2011, gutman_frozen_2012, gutman_slow_2012}, although without explicitly adopting the terminology of exceptional points. Dispersion relations exhibiting stationary inflection points have also been reported in Refs.~\cite{sumetsky2005uniform,mori2005wideband, baba2007slow,kawasaki2007experimental, scheuer2011serpentine, scheuer2011optical}, without identifying them as EPDs. The order of an EPD is determined by the number of coalescing eigenmodes \cite{nada_theory_2017} and plays a fundamental role in determining the electromagnetic response of finite-length structures \cite{nada_theory_2017,nadaDesignModifiedCoupled2023}. However, operation near an EPD can also exhibit pronounced sensitivity to parameter perturbations, making the fabrication of photonic integrated circuits supporting higher-order EPDs particularly challenging. Consequently, experimental demonstrations of such degeneracies in optical structures remain relatively limited, with notable examples including the degenerate band edge in \cite{burr_degenerate_2013} and the SIP in \cite{Paul2024experimental,furman2025dispersionEng,furman2025experimentalSIP}.

\begin{figure}[ht]
    \centering
    \begin{subfigure}[t]{\linewidth}
        \centering
        \includegraphics[width=\textwidth]{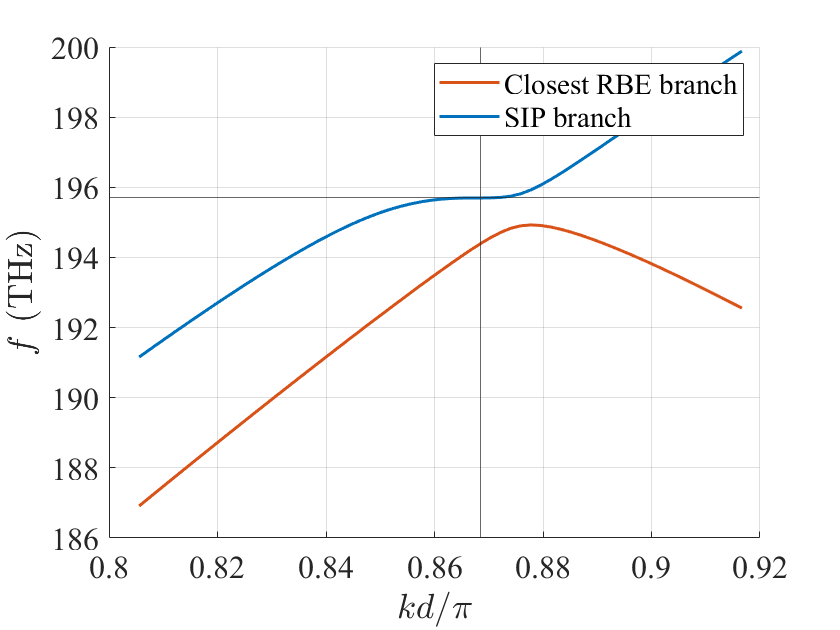}
        \caption{}
        \label{fig:DispDiagrSIP}
    \end{subfigure}
    \hfill
    \begin{subfigure}[t]{\linewidth}
        \centering
        \includegraphics[width=\textwidth]{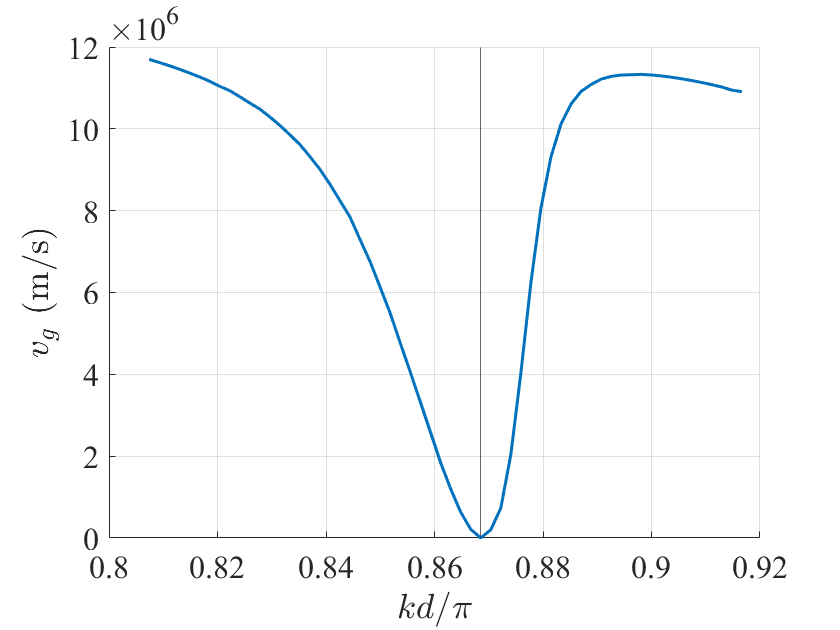}
        \caption{}
        \label{fig:DispDiagrvg}
    \end{subfigure}
    \caption{(a) Modal dispersion diagram of the TWG2RA waveguide showing the SIP-supporting branch (blue) and the branch containing the nearest regular band edge (RBE, orange). Only branches with purely real wavenumber are shown. (b) Group velocity $v_g=\partial\omega/\partial k$ as a function of the Bloch wavenumber $k$ near the SIP. The group velocity vanishes at $k=k_{\mathrm{S}}$ and exhibits a quadratic dependence on the wavenumber detuning, $v_g\propto(k-k_{\mathrm{S}})^2$, consistent with the cubic dispersion relation $\omega-\omega_{\mathrm{S}}\propto(k-k_{\mathrm{S}})^3$ characteristic of a third-order SIP.}
    \label{fig:DispersionFeatures}
\end{figure}

One of the most remarkable properties of the SIP is the coalescence of three Bloch eigenmodes, involving propagating and evanescent modes \cite{nada_theory_2017,parareda_frozen_2022,nadaDesignModifiedCoupled2023}, which gives rise to the frozen mode regime characterized by vanishing group velocity and strong field enhancement \cite{figotin_oblique_2003,furman_impact_2025}. The SIP occurs at a point $(k_{\mathrm{S}},\omega_{\mathrm{S}})$ in the Bloch dispersion relation where the group velocity $v_g=\partial\omega/\partial k$ vanishes (Fig.~\ref{fig:DispersionFeatures}(a)) together with its first derivative with respect to the Bloch wavenumber, i.e., $\partial\omega/\partial k=0$, and $\partial v_g/\partial k=\partial^2\omega/\partial k^2=0$, as shown in Fig.~\ref{fig:DispersionFeatures}(b), while $\partial^3\omega/\partial k^3\neq 0$. Consequently, the dispersion relation near the SIP exhibits a cubic dependence, $\omega-\omega_{\mathrm{S}}\propto(k-k_{\mathrm{S}})^3$, and the group velocity approaches zero quadratically as $k$ approaches $k_S$. In the frozen mode regime, the interference between propagating and evanescent Bloch modes allows the electromagnetic field to satisfy the boundary conditions at the entrance of the structure while developing a strongly enhanced amplitude inside the periodic waveguide \cite{figotin_oblique_2003,figotin_slow_2011}. In comparison, a second-order EPD associated with a regular band edge (RBE) involves the coalescence of two counter-propagating Bloch modes and leads to the formation of a conventional standing-wave resonance \cite{nada_theory_2017}. Higher-order EPDs involve a more complex interaction among propagating and evanescent modes \cite{figotin_gigantic_2005, figotin_slow_2006, nada_theory_2017}. The SIP represents the lowest-order EPD supporting the frozen mode regime and is characterized by the coalescence of three Bloch eigenmodes and their corresponding eigenvectors \cite{nada_theory_2017, parareda_frozen_2022}.

Figure~\ref{fig:DispersionFeatures}(a) shows the real branches of the Bloch dispersion diagram of an SIP-supporting waveguide, while Fig.~\ref{fig:DispersionFeatures}(b) illustrates the corresponding group velocity $v_g$ as a function of the Bloch wavenumber $k$ near the SIP. The group velocity vanishes at the SIP wavenumber $k_{\mathrm{S}}$, where it exhibits a local minimum. Although only the propagating branches are shown in Fig.~\ref{fig:DispersionFeatures}(a), the SIP involves the coalescence of three Bloch eigenmodes, including evanescent modes that are not represented in the figure (they are visible in \cite{gutman_slow_2012,nada_theory_2017,parareda_frozen_2022, furman2025dispersionEng}, for example). Unlike an RBE, which involves two counter-propagating modes, a third-order EPD supporting the frozen mode regime requires the coupling of at least three Bloch modes. In a reciprocal periodic waveguide, this requires at least three coupled modal channels in each propagation direction, including both propagating and evanescent solutions, as discussed in \cite{figotin_slow_2011, gutman_slow_2012, nada_theory_2017, parareda_frozen_2022, furman2025dispersionEng}.

Operating near an SIP offers several advantages over an RBE. In particular, the SIP supports the frozen mode regime without being adjacent to a bandgap and, as a third-order EPD, is generally less sensitive to perturbations than higher-order EPDs. Although the group velocity dispersion near the SIP is strongly frequency-dependent and asymmetric, the resulting signal distortion can, in principle, be compensated if the spectral response of the waveguide is properly characterized. Owing to these distinctive dispersion properties, SIP-based structures have been proposed for achieving large true-time delays \cite{baba2007slow,paul_frozen_2021,paul2022photonic}. They have also attracted considerable interest for developing unconventional lasing regimes \cite{ramezani2014unidirectional, parareda_lasing_2023,furman_impact_2025,zamir2023low} and nonlinear optical isolators.

\begin{figure}[ht]
    \centering

    \begin{subfigure}[t]{0.75\linewidth}
        \centering
        \includegraphics[width=\textwidth]{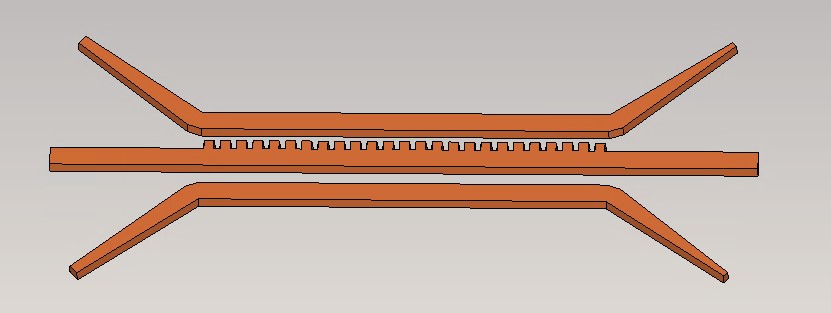}
        \caption{}
        \label{fig:SIPWaveguides_b}
    \end{subfigure}

    \begin{subfigure}[t]{0.75\linewidth}
        \centering
        \includegraphics[width=\textwidth]{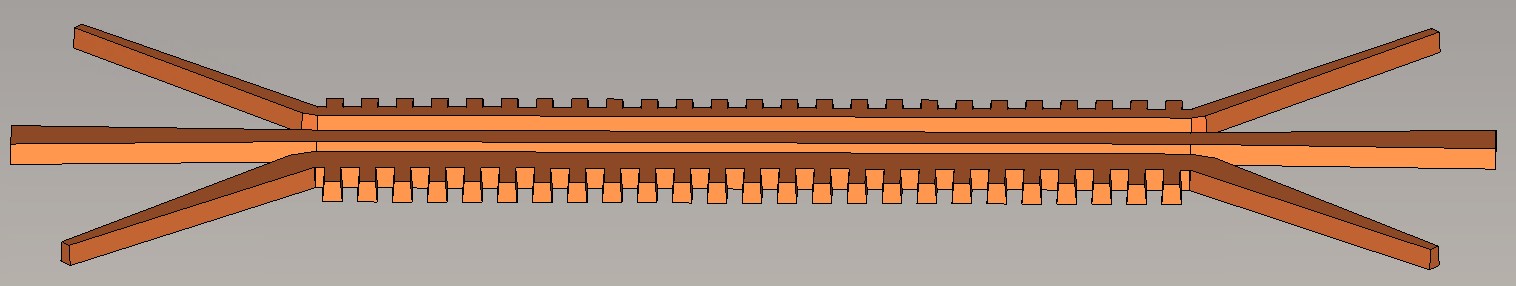}
        \caption{}
        \label{fig:SIPWaveguides_a}
    \end{subfigure}

    \vspace{3mm}

    \vspace{3mm}

    \begin{subfigure}[t]{0.75\linewidth}
        \centering
        \begin{minipage}[t]{0.8\linewidth}
            \centering
            \includegraphics[width=\textwidth]{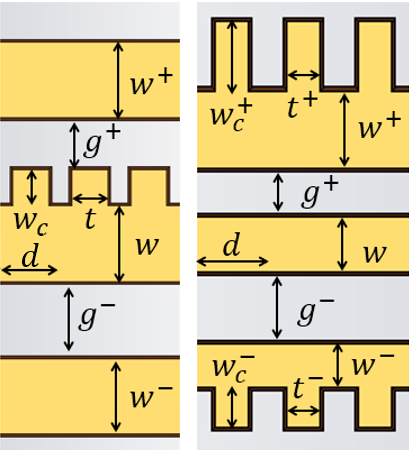}
        \end{minipage}%
        
        \caption{}
        \label{fig:SIPWaveguides_c}
    \end{subfigure}
    \caption{Schematic representations of the two proposed waveguide configurations: (a) TWG1, consisting of a central waveguide with lateral gratings evanescently coupled to two straight waveguides; (b) TWG2, consisting of a central straight waveguide evanescently coupled to two waveguides with lateral gratings; and (c) top views of the TWG1 and TWG2 configurations, highlighting their respective periodic unit cells of length $d$.}
    \label{fig:SIPWaveguides}
\end{figure}

We introduce a systematic methodology for designing periodic silicon photonic waveguides supporting an SIP through dispersion engineering. The proposed structures consist of three coupled single-mode waveguides incorporating lateral gratings, providing a fabrication-compatible approach to realizing the frozen mode regime. Two configurations are considered:  a central waveguide with lateral gratings coupled to two adjacent straight waveguides (TWG1), and a central straight waveguide coupled to two laterally corrugated waveguides (TWG2), as illustrated in Fig.~\ref{fig:SIPWaveguides}(a) and (b), respectively. In both configurations, the SIP is achieved through the synchronization and coalescence of three Bloch eigenmodes, involving one propagating and two evanescent modes. We develop and analyze three designs, two based on TWG2 and one on TWG1, using full-wave eigenmode simulations to optimize their dispersion characteristics. 

To the best of the authors' knowledge, the realization of SIP-supporting photonic integrated waveguides based on three coupled single-mode waveguides incorporating lateral gratings has not been previously reported. This configuration provides an alternative to existing SIP-supporting photonic structures, combining a relatively simple geometry with compatibility with established silicon photonic fabrication processes. We further investigate the robustness of the proposed designs by examining the effects of geometric perturbations on the dispersion characteristics and assessing the numerical convergence of the results under different simulation mesh resolutions. Following the design optimization and wafer fabrication, we experimentally characterize the transmission spectra of finite-length periodic waveguides and compare the measured spectral features with those predicted by full-wave simulations. Finally, we investigate the group-delay characteristics numerically, demonstrating substantial delay enhancement near the SIP frequency.

The paper is organized as follows: The methodology to design SIP-supporting three-way waveguides is described in Section~\ref{ch:waveguideDesign}. In Section~\ref{ch:tapeoutPrep}, we highlight the expected differences between design and fabrication, leading to a redesign that is more consistent with fabrication methods. The experimental features of finite-length SIP waveguides are discussed in Section~\ref{ch:Measurement}. The results are summarized in Section~\ref{ch:Conclusion}.

\section{Process steps to design an SIP}
\label{ch:waveguideDesign}

The {\em first step} in designing an SIP-supporting PIC consists of coupling a waveguide incorporating a lateral grating to a straight waveguide. Their geometrical parameters are selected such that the interaction between counter-propagating modes opens a bandgap near a target point $(k_{\mathrm{S}}, f_{\mathrm{S}})$ in the Brillouin zone. A third waveguide, supporting an additional optical mode, is then placed adjacent to the coupled pair. As the separation between the waveguides decreases, the resulting three-way modal coupling modifies the dispersion relation, producing local maxima and minima near the target frequency $f_{\mathrm{S}}$. By progressively tuning the waveguide dimensions and coupling strengths, these extrema can be brought together to form a stationary inflection point at the desired location in the dispersion diagram.

\begin{figure*}[t]
    \centering
    \begin{subfigure}[t]{0.5\textwidth}
        \centering
        \includegraphics[width=\textwidth]{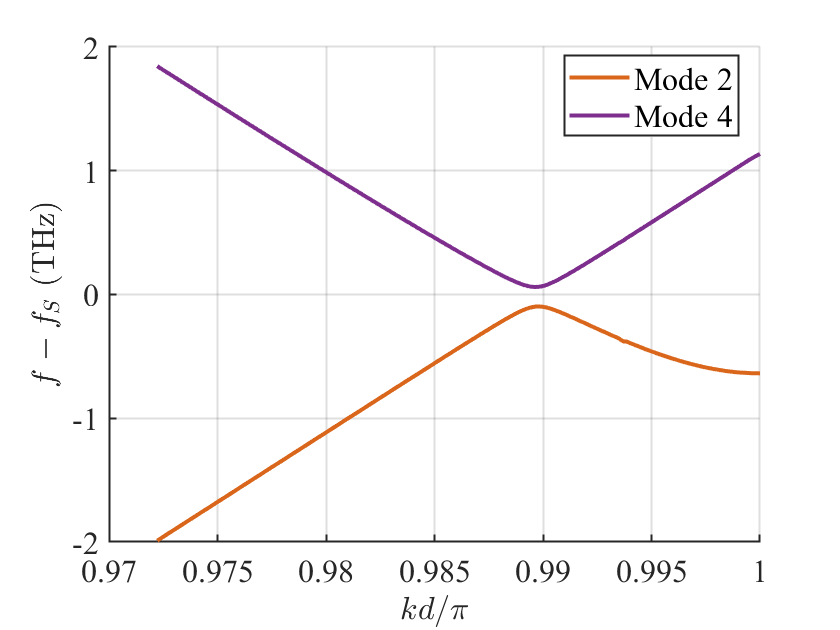}
        \caption{}
        \label{fig:SIPsearch_a}
    \end{subfigure}\hfill
    \begin{subfigure}[t]{0.5\textwidth}
        \centering
        \includegraphics[width=\textwidth]{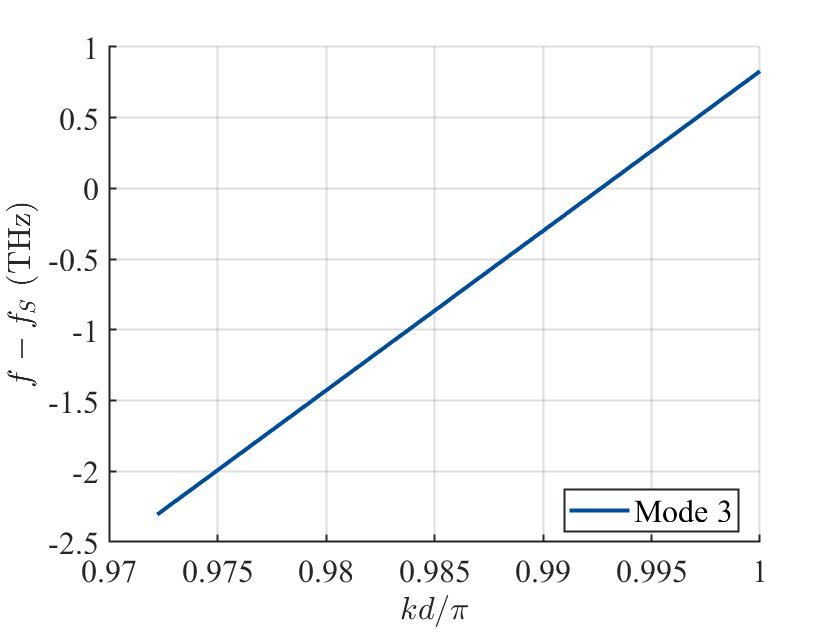}
        \caption{}
        \label{fig:SIPsearch_b}
    \end{subfigure}
    \hfill
    \begin{subfigure}[t]{0.5\textwidth}
        \centering
        \includegraphics[width=\textwidth]{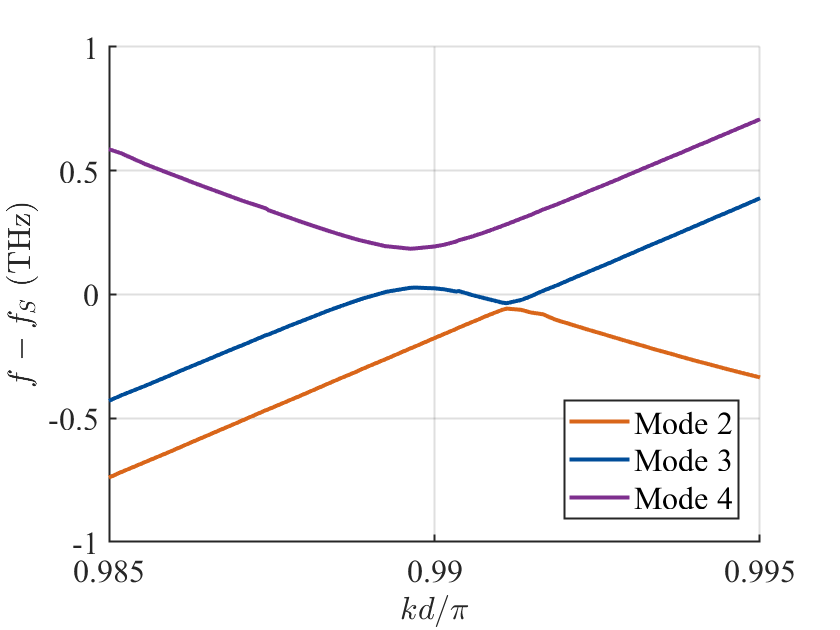}
        \caption{}
        \label{fig:SIPsearch_c}
    \end{subfigure}\hfill
    \begin{subfigure}[t]{0.5\textwidth}
        \centering
        \includegraphics[width=\textwidth]{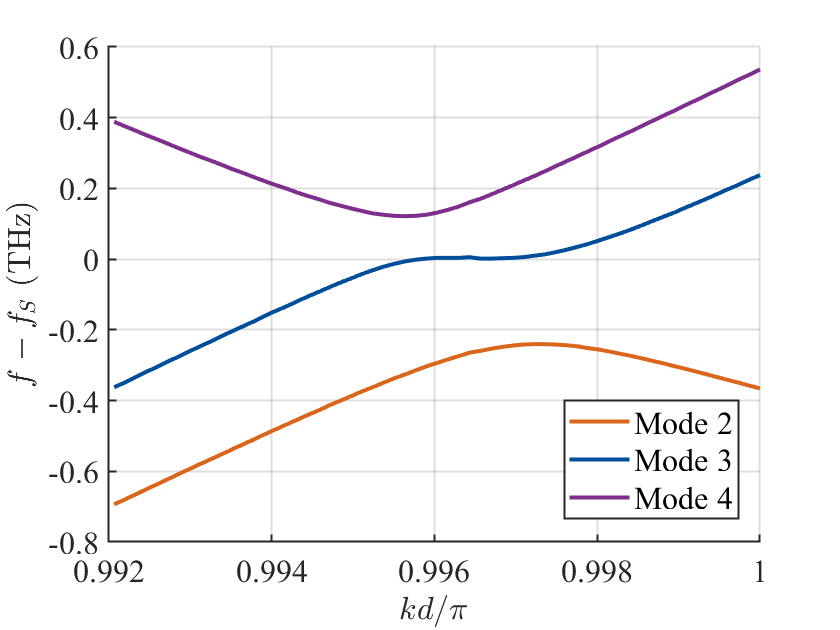}
        \caption{}
        \label{fig:SIPsearch_d}
    \end{subfigure}
    \caption{Design steps for creating a stationary inflection point (SIP)-based frozen mode photonic integrated circuit (PIC) using the TWG1 configuration. (a) Dispersion diagram of a grating coupled to a straight waveguide, with parameters selected to open a bandgap near the target point $(k_{\mathrm{S}}, f_{\mathrm{S}})$. (b) Dispersion diagram of an isolated straight waveguide showing a mode near the same target region. (c) Dispersion diagram of the TWG1 configuration, formed by adding the additional waveguide from (b) to the grating–straight waveguide pair in (a). The resulting structure produces the blue dispersion curve with a local maximum and minimum near the target point $(k_{\mathrm{S}}, f_{\mathrm{S}})$. (d) Optimized TWG1 dispersion diagram after adjusting waveguide gap sizes, where a stationary inflection point (SIP) emerges at the target $(k_{\mathrm{S}}, f_{\mathrm{S}})$ location (blue curve).}
    \label{fig:SIPsearch}
\end{figure*}

Figure~\ref{fig:SIPsearch} illustrates the SIP design process using the TWG1 configuration as an example. Panels (a) and (b) show the individual dispersion diagrams of the (a) grating–straight waveguide pair, and (b) the individual third waveguide, respectively. In (c), the lightly coupled three-way structure shows local maxima and minima near $(k_{\mathrm{S}}, f_{\mathrm{S}})$. In (d), after optimizing parameters such as waveguide widths and gaps, an SIP is achieved, identified by a stationary inflection in the Bloch dispersion diagram with vanishing group velocity at $(k_{\mathrm{S}}, f_{\mathrm{S}})$. Similar dispersion diagrams are shown in \cite{gutman_frozen_2012} depending on the value of a waveguide parameter.

We designed two TWG2 waveguides and one TWG1 waveguide, denoted as TWG2A, TWG2B, and TWG1A, respectively, supporting an SIP at a frequency $f_{\mathrm{S}}$ in the range $[192,195]$ THz. The geometrical parameters and dispersion characteristics of these designs are summarized in Table~\ref{table:SIPdesign}. An important figure of merit is the frequency separation $\Delta f_{SR}$ between the SIP and the nearest RBE. A large separation is desirable to spectrally isolate the frozen mode regime from neighboring RBE effects, particularly in long waveguides where resonances associated with different dispersion features may become closely spaced. In general, reducing the unit-cell period $d$ can increase $\Delta f_{SR}$, consistent with the increase in the characteristic free spectral range $\Delta f_{FSR}=c/(n_w d)$ \cite{scheuer2011serpentine}. Previously reported photonic waveguides supporting an SIP exhibit estimated values of $\Delta f_{SR}$ (in GHz) of 344 \cite{paul_frozen_2021}, 60 \cite{nada_theory_2017}, 31 \cite{parareda_frozen_2022}, 80 \cite{parareda_lasing_2023}, 306 \cite{furman_frozen_2023}, 186 \cite{zamir2023low}, and 310 \cite{furman2025experimentalSIP}. The proposed designs achieve comparatively large SIP--RBE separations, particularly the TWG2RA configuration discussed later, for which $\Delta f_{SR}=1000$ GHz. Although the separation depends on several geometrical parameters and the resulting modal coupling, the relatively small unit-cell period of the proposed lateral-grating waveguides contributes to the large spectral separation between the SIP and nearby RBEs.
\begin{table}[ht]
\caption{SIP-supporting waveguide designs}
\centering
\begin{tabularx}{\linewidth}{|c|X|X|X|}
\hline
\textbf{Params} (nm) & \textbf{TWG2A} & \textbf{TWG2B} & \textbf{TWG1A} \\
\hline
$w$ & 310 & 300 & 450  \\
$w^+$ & 250 & 250 & 480  \\
$w^-$ & 420 & 325 & 477  \\
$g^+$ & 325 & 325 & 200  \\
$g^-$ & 200 & 200 & 291  \\
$w_c^+$ & 200 & 200 & 200 \\
$w_c^-$ & 380 & 466 & -  \\
$t^+$ & 200 & 200 & 200  \\
$t^-$ & 200 & 200 & -  \\
$d$ & 350 & 370 & 325  \\
\hline
$f_{\mathrm{S}}$ (THz) & 192.28 & 193.13 & 192.95 \\
$k_{\mathrm{S}} d/\pi$ & 0.983 & 0.97 & 0.87 \\
$\Delta f_{SR}$ (GHz) & 540 & 565 & 80 \\
\hline
\end{tabularx}
\label{table:SIPdesign}
\end{table}

The modal dispersion relations were calculated using the full-wave eigenmode solver of CST Studio Suite, based on the finite element method (FEM). Phase-shifted periodic boundary conditions were imposed at two transverse cross-sections separated by one unit-cell period, $d$, to determine the Bloch eigenmodes and their corresponding dispersion characteristics. In the transverse directions, perfect electric conductor (PEC) boundary conditions were applied sufficiently far from the waveguide cores, with a thick silicon dioxide cladding region ensuring that the evanescent fields decay to negligible amplitudes before reaching the computational boundaries. Material dispersion was neglected over the frequency range of interest \cite{dattner_analysis_2011}, and the refractive indices of the silicon core and silicon dioxide cladding were assumed to be frequency-independent, with values $n_{\mathrm{Si}}=3.478$ and $n_{\mathrm{SiO_2}}=1.444$, respectively.

\section{Wafer tapeout preparation}
\label{ch:tapeoutPrep}
SIP-based devices are generally sensitive to geometrical perturbations, as is characteristic of systems operating near EPDs \cite{kato_perturbation_1966, figotin_frozen_2006, nada_theory_2017}. Even small deviations from the nominal geometry can modify the dispersion characteristics and disrupt the coalescence of Bloch eigenmodes required to establish an SIP. Nevertheless, a relatively flat dispersion profile, and hence the potential for enhanced group delay, may persist even when the exact SIP condition is no longer satisfied. To account for realistic fabrication conditions and restore the SIP, we therefore redesign the proposed waveguides by incorporating fabrication constraints and anticipated geometrical deviations. An important fabrication constraint is the minimum achievable feature size, typically on the order of 100 nm.

\begin{figure}[ht]
    \centering
    \begin{subfigure}[t]{0.95\linewidth}
        \centering
        \includegraphics[width=\textwidth]{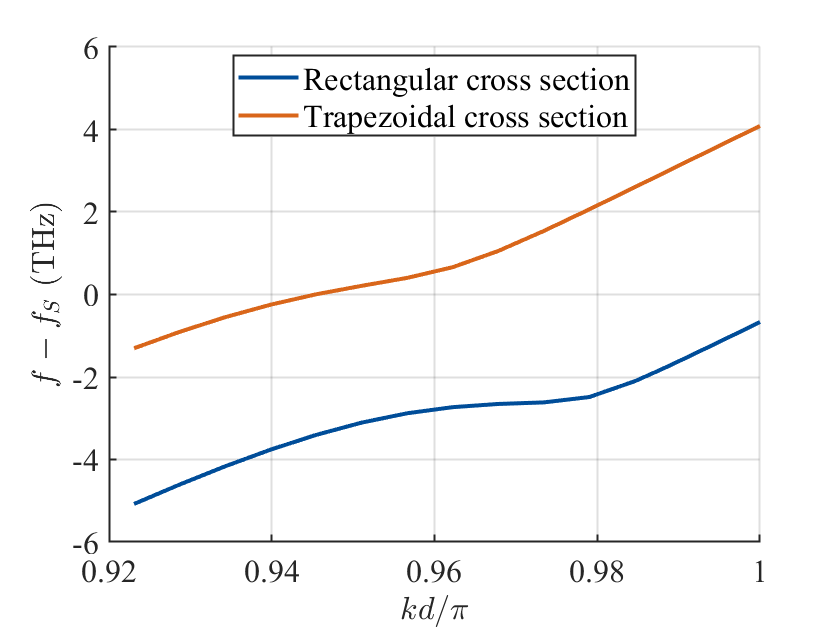}
        \label{fig:Rhomboid_c}
    \end{subfigure}
    \caption{Dispersion diagram of the SIP-supporting TWG2A waveguide with rectangular and trapezoidal cross-sections. The inclined sidewalls of the trapezoidal geometry modify the modal characteristics and slightly disrupt the SIP condition, which can be restored through fine-tuning of the waveguide parameters.}
    \label{fig:Rhomboid}
\end{figure}

Another common fabrication-induced deviation is the departure of the waveguide cross-section from its nominal rectangular shape. In particular, the etching process typically produces slightly inclined sidewalls, resulting in a trapezoidal cross-section with a narrower top surface. This geometrical deviation modifies the modal field distributions and the coupling between adjacent waveguides, potentially affecting the SIP condition. The impact of the sidewall inclination is illustrated in Figure~\ref{fig:Rhomboid}, which compares the dispersion diagrams of the TWG2A waveguide with ideal rectangular and realistic trapezoidal cross-sections. Introducing a realistic sidewall angle (i.e., by decreasing the top dimensions based on tilting all walls by a few degrees $\theta_\delta$ from the $90^{\degree}$) slightly perturbs the dispersion near the SIP, disrupting the exact coalescence of the three Bloch eigenmodes. Nevertheless, the dispersion remains relatively flat, indicating that the slow-light characteristics are largely preserved. Fine-tuning of the geometrical parameters is therefore required to restore the exact SIP condition. Waveguides are nominally $220$ nm wide and the top width is adjusted to account for the realistic sidewall angle $\theta_\delta$ \cite{fahrenkopfAIMmpw2019}. Despite the geometrical perturbation introduced by the inclined sidewalls, the TWG2A waveguide retains a relatively flat dispersion profile near the original SIP frequency, indicating that its slow-light characteristics are reasonably robust against deviations from the nominal geometry. Although the exact SIP condition is disrupted, the low group velocity near the original SIP frequency is expected to produce enhanced group delay. Nevertheless, restoring the coalescence of the three Bloch eigenmodes, and hence the exact SIP condition, requires further optimization of the geometrical parameters, as described next.

The optimized geometrical parameters of the redesigned waveguides, which support an SIP while accounting for realistic trapezoidal cross-sections, are summarized in Table~\ref{table:Redesign}. All reported transverse dimensions refer to the bottom of the waveguide cross-sections. The hyphens in the TWG1RA column indicate parameters that are not applicable to this configuration, since TWG1RA incorporates only one waveguide with lateral gratings, whereas the TWG2R A and B configurations incorporate two.


\begin{figure}
    \centering
 \includegraphics[width=0.95\linewidth]{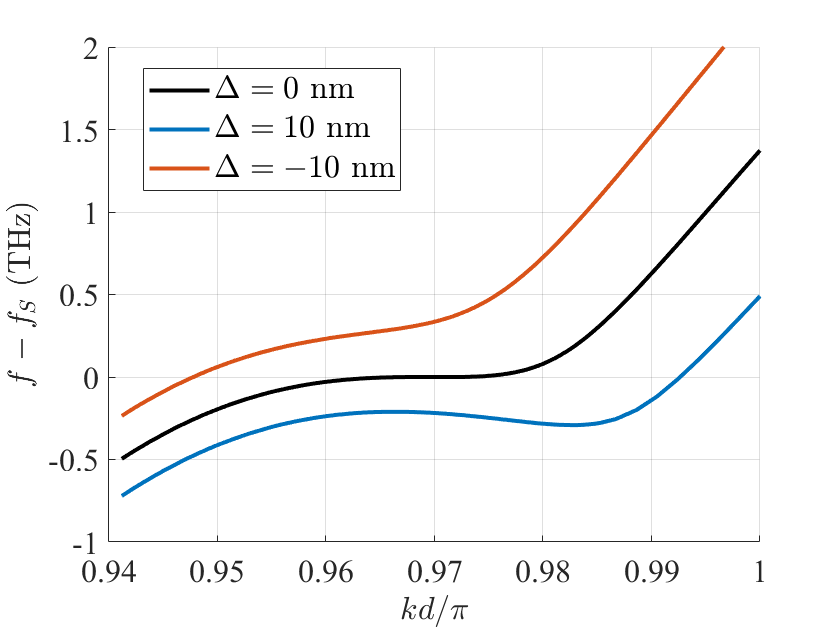}
    \caption{Dispersion diagram of the TWG2RB waveguide for the nominal geometry (black) and for waveguide widths uniformly increased (blue) or decreased (red) by $\Delta=10$~nm. These geometrical perturbations disrupt the exact SIP condition but preserve a relatively flat dispersion profile near the original SIP frequency. The results highlight the sensitivity of Bloch mode coalescence to fabrication-induced variations, while indicating that slow-light characteristics may persist even when the exact SIP condition is not fully satisfied.}
    \label{fig:ToleranceFig}
\end{figure}
Preparing frozen-mode PICs for wafer tapeout requires a fabrication tolerance analysis to assess how geometrical deviations may affect device performance. While previous analytical approaches have modeled fabrication tolerances through variations in the effective refractive index \cite{furman_impact_2025}, here we directly investigate numerically the effects of geometrical perturbations on the Bloch dispersion characteristics, with particular emphasis on the preservation or disruption of the mode coalescence required to establish an SIP.

\begin{table}[ht]
\caption{Realistic SIP-supporting waveguide designs, assuming tilted walls}
\centering
\begin{tabularx}{\linewidth}{|c|X|X|X|}
\hline
\textbf{Params} (nm) & \textbf{TWG2RA} & \textbf{TWG2RB} & \textbf{TWG1RA} \\
\hline
$w$ & 310 & 300 & 450  \\
$w^+$ & 250 & 250 & 480  \\
$w^-$ & 420 & 325 & \textbf{477}  \\
$g^+$ & \textbf{319} & 325 & 200  \\
$g^-$ & 200 & 200 & \textbf{305}  \\
$w_c^+$ & 200 & 200 & 200  \\
$w_c^-$ & \textbf{340} & \textbf{436} & -  \\
$t^+$ & 200 & 200 & 200  \\
$t^-$ & 200 & 200 & -  \\
$d$ & \textbf{355} & 370 & 325  \\
\hline
$f_{\mathrm{S}}$ (THz) & 195.755 & 197.58 & 194.16  \\
$k_{\mathrm{S}} d/\pi$ & 0.867 & 0.91 & 0.96  \\
$\Delta f_{SR}$ (GHz) & 1000 & 664.5 & 127.6  \\
\hline
\end{tabularx}
\label{table:Redesign}
\end{table}

Fabrication tolerances are generally around $2-5$ nm \cite{fahrenkopfAIMmpw2019, furman_impact_2025} which are either added or subtracted from the nominal geometric parameters. Figure~\ref{fig:ToleranceFig} shows the dispersion diagram of the TWG2RB design. The nominal case, in black, and perturbed cases with $\Delta = \pm 10$ nm applied collectively to the waveguide widths $w$, $w^+$, and $w^-$ in blue and red, respectively. 
The geometrical width perturbations $\Delta$ modify the dispersion near the SIP, giving rise to different types of tilted inflection points (TIPs) \cite{marosi_three_2022}. As shown in Figure~\ref{fig:ToleranceFig}, uniformly decreasing the waveguide widths by $\Delta=-10$ nm produces a relatively small deviation from the nominal SIP dispersion. The resulting red curve exhibits a smooth-TIP, characterized by a nearly stationary inflection point with a small but nonzero group velocity. In contrast, increasing the widths by $\Delta=+10$ nm produces the blue curve, which exhibits a local maximum and a local minimum near the original SIP frequency. This behavior corresponds to an alternating-TIP, characterized by a change in the sign of the group velocity between the two stationary points. Similar dispersion profiles arise during the SIP design procedure described in Section~\ref{ch:waveguideDesign} and were also investigated in \cite{gutman_frozen_2012}, where the transition between different dispersion regimes was demonstrated by fine-tuning a geometrical parameter.

Comparable behavior is observed when perturbing the corrugation widths, coupling gaps, or waveguide thicknesses in the TWG2RA configuration (not shown), as well as in the TWG2RB and TWG1RA designs. These results reveal an important distinction between the sensitivity of the exact SIP condition and the persistence of slow-light characteristics under geometrical perturbations. Although relatively small dimensional changes can noticeably alter the dispersion and disrupt the exact coalescence of the Bloch eigenmodes, the perturbed structures retain relatively flat dispersion profiles with regions of low group velocity, potentially supporting enhanced group delay. The response is also asymmetric with respect to the sign of the perturbation, as illustrated by the qualitatively different dispersion profiles obtained for $\Delta=\pm10$ nm.

Importantly, the considered perturbations of $\pm10$ nm exceed the expected fabrication tolerances of approximately 2--5 nm. Nevertheless, the simulated dispersion profiles retain low-group-velocity regions near the nominal SIP frequency. This suggests that the proposed waveguides can preserve favorable slow-light characteristics under moderate geometrical deviations, even when the exact SIP condition is disrupted. Thus, while accurate fabrication remains essential for achieving the intended third-order mode coalescence, the frozen-mode-related dispersion characteristics exhibit a degree of tolerance to fabrication-induced perturbations.
\begin{figure}[ht]
    \centering
    \begin{subfigure}[t]{0.95\linewidth}
        \centering
        \includegraphics[width=\textwidth]{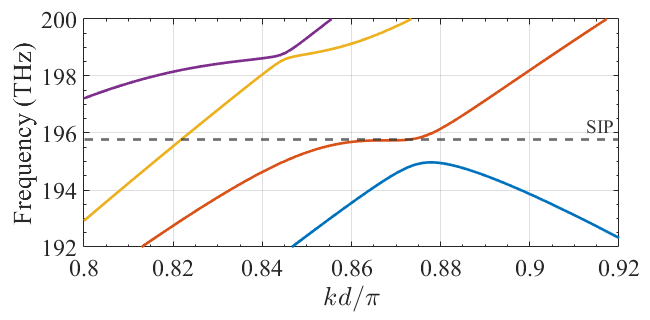}
    \end{subfigure}
    \caption{Dispersion diagram of the SIP-supporting TWG2RA waveguide with a realistic trapezoidal cross-section. The SIP-supporting branch (orange) and the three neighboring branches with purely real-valued Bloch wavenumbers are shown. The nearest RBE occurs at a frequency $1$ THz below the SIP, corresponding to an SIP--RBE separation of $\Delta f_{SR}=1000$ GHz.}
    \label{fig:DispTWG2RA}
\end{figure}
Figure~\ref{fig:DispTWG2RA} shows the modal dispersion diagram of the redesigned TWG2RA waveguide with a realistic trapezoidal cross-section, exhibiting an SIP at $f_{\mathrm{S}}=195.755$ THz. The figure also shows the upper and lower dispersion branches and their corresponding critical points. In particular, the nearest RBE, located on the blue branch, occurs approximately $1$ THz below the SIP on the orange branch, yielding a large SIP--RBE frequency separation of $\Delta f_{SR}=1000$ GHz. This substantial spectral separation is a distinguishing feature of the TWG2RA design. In the following section, we investigate the transmission characteristics and group delay of finite-length waveguides based on this configuration.

\section{Finite-Length Simulated and Measurement Results}
\label{ch:Measurement}

\begin{figure}
    \centering
    \begin{subfigure}[t]{\linewidth}
       \centering
       \includegraphics[width=\textwidth]{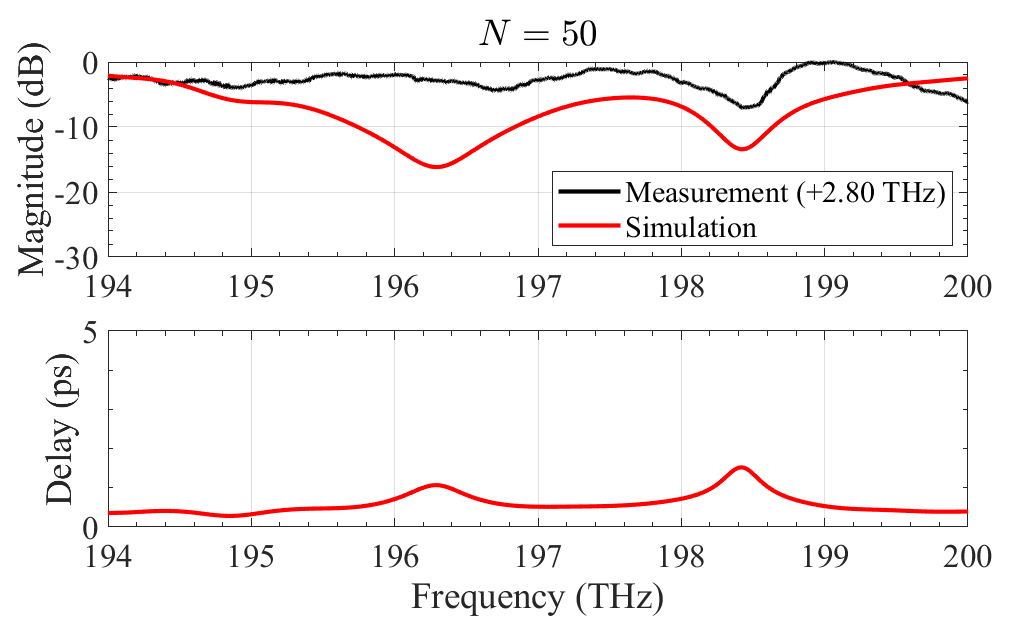}
       \caption{}
    \end{subfigure}
    \hfill
    \begin{subfigure}[t]{\linewidth}
       \centering
       \includegraphics[width=\textwidth]{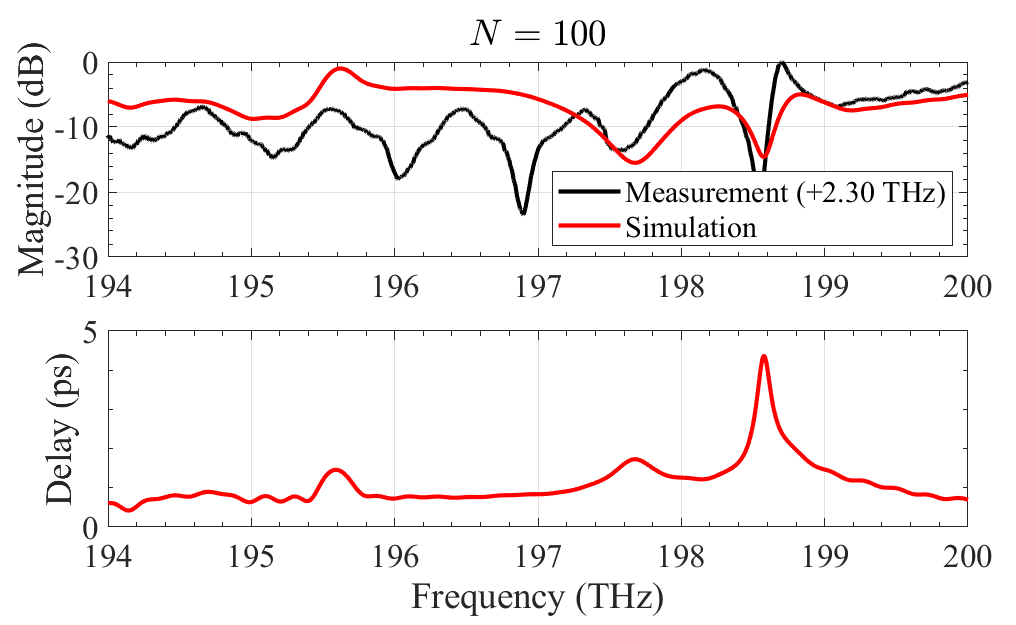}
       \caption{}
    \end{subfigure}
    \hfill
    \begin{subfigure}[t]{\linewidth}
       \centering
       \includegraphics[width=\textwidth]{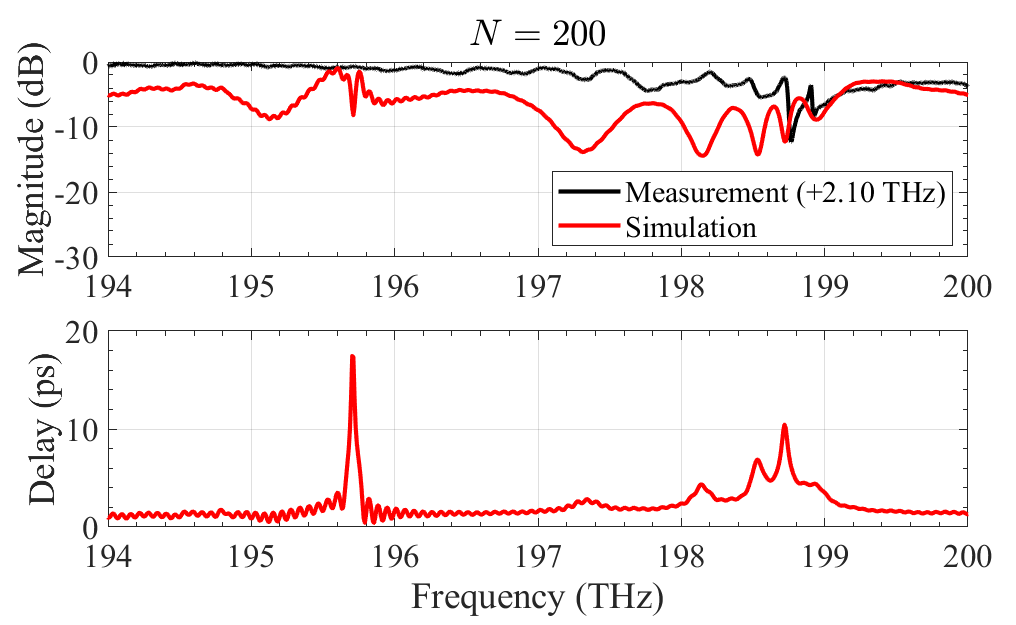}
       \caption{}
    \end{subfigure}
    \hfill
    \caption{Transmission spectra and group delay of the TWG2RA waveguide with (a) $N=50$, (b) $N=100$, and (c) $N=200$ unit cells. Measured and simulated transmission spectra are compared, while the group delay is obtained from simulations only. The measured spectra are shifted in frequency by the amounts indicated in the figure to facilitate comparison with the simulations. As $N$ increases, the simulated group delay exhibits a progressively stronger enhancement near the SIP frequency, $f_{\mathrm{S}}=195.755$ THz, compared with the delay associated with spectral features around $198.6$ THz.}
    \label{fig:SpectralResponse}
\end{figure}
We investigate the transmission characteristics and group delay of finite-length TWG2RA waveguides through numerical simulations and compare the simulated transmission spectra with experimental measurements. The devices were fabricated by the American Institute for Manufacturing Integrated Photonics (AIM Photonics) through its multi-project wafer (MPW) fabrication process \cite{fahrenkopfAIMmpw2019}.
 We focus on the wafer-ready SIP-TWG2RA design (parameters given in Table~\ref{table:Redesign}) with lengths of $N=50$, $N=100$, and $N=200$ unit cells. 
 The fabricated devices follow the same geometrical design used in the full-wave simulations where we also consider the sidewall angling. As shown in Fig.~\ref{fig:SIPWaveguides_a}, at both ends of the periodic structure, the terminations of the outer two waveguides are angled $40\degree$ away from the center waveguide with a taper length of $4\;\mathrm{\upmu m}$ and a minimum width of $200\;\mathrm{nm}$, which is the same choice for the terminations adopted in \cite{furman2025experimentalSIP}. These tapered terminations have two purposes: (i) they progressively  reduce the coupling between the outer and central waveguides and (ii) they strongly reduce individual-waveguide reflections. Hence, they provide a transition from the three coupled waveguides in the periodic section to the single central waveguide used for optical excitation and collection. The termination design is particularly important for testing SIP-supporting structures because the boundary conditions strongly influence the excitation and interference of the Bloch modes, reducing conventional Fabry-P\'erot resonances and highlighting the effect of the SIP that tents to be naturally mismatched to any combination of the three-waveguide individual terminations. Consequently, 
the tapered terminations adopted here are intended to limit unwanted reflections and facilitate the identification of spectral transmission features like the group delay associated with the frozen mode regime   as  discussed by Furman et al. \cite{furman2025experimentalSIP}.

We measured the magnitude of the transfer function of these edge-coupled devices using a high power superluminescent diode~(SLD) paired with an optical spectrum analyzer~(OSA). The SLD has a spectrum centered around $\lambda = 1550\;\mathrm{nm}$~($\approx 193.4\;\mathrm{THz}$) with a significant, $\approx$ 20 dB, signal drop-off at wavelengths smaller than $\approx 1515\;\mathrm{nm}$~($\approx 197.9\;\mathrm{THz}$). The simulation data is presented as-is and the measured data is normalized to the SLD spectrum and has its maximum magnitude in the shown frequency range set to $0\;\mathrm{dB}$. The SLD/OSA measurement setup well characterizes the spectral response of the devices yet does not have the capability to measure the group delay.

Figure~7 compares the measured (black) and simulated (red) transmission spectra of the TWG2RA waveguides with $N=50$, $100$, and $200$ unit cells. The corresponding simulated group delays are also shown; group delay was not measured experimentally. To facilitate comparison, the measured spectra are shifted in frequency by the amounts indicated in the figure legends. These shifts are determined by maximizing the cross-correlation between the measured and simulated transmission spectra over the frequency range from $196$ to $200$ THz, as detailed in \cite{furman2025experimentalSIP}. The required frequency shifts are comparable to those observed for other devices and waveguide designs fabricated on the same wafer, suggesting that they are consistent with typical fabrication-induced variations.

We calculate the group delay based on the phase of the transfer function $\angle T_f(\omega)$ as
\begin{equation}
    \tau_g=-\frac{\partial \angle T_f(\omega)}{\partial \omega},
    \label{eq:GroupDelay}
\end{equation}
where $\omega=2\pi f=2\pi c_0 / \lambda$ is the angular frequency and $c_0$ is the speed of light in vacuum. As the phase changes more rapidly over some frequency, we expect to see an increase in the group delay.

Given a waveguide width of $w = 480\;\mathrm{nm}$ with the modal index calculated from simulations of $n_w = 2.378$, the delay per length in a straight uniform waveguide of this width is $\approx 7.9\;\mathrm{fs/\upmu m}$. Using the period of $d=355\;\mathrm{nm}$ from Table~\ref{table:Redesign}, the total length of the $N = 50$, $N = 100$ and $N=200$ devices are $17.75\;\mathrm{\upmu m}$,  $35.50\;\mathrm{\upmu m}$, and $71.0 \;\mathrm{\upmu m}$, respectively. This corresponds to a total delay in an equivalent straight waveguide of the same length with $w = 480\;\mathrm{nm}$ of $0.140\;\mathrm{ps}$,  $0.280\;\mathrm{ps}$, and $0.560\;\mathrm{ps}$. From simulations, we see delays significantly larger than these values around features of interest. In particular, around the SIP frequency, we observe a very significant increase in delay for $N=200$. For this length, the group delay is $18\;\mathrm{ps}$ that is $\approx 32$ times larger than the $0.560\;\mathrm{ps}$ delay of a uniform straight waveguide of the same length. 
While this delay peak is also seen for $N=100$, for this length the delay peak at nearby spectral features at  $198.6\;\mathrm{THz}$ is more significant. Only when $N$ becomes sufficiently large, as for $N=200$, do we observe the delay around the SIP to be quite large and even dominate with respect to that at higher frequencies (compare the scale of the group delay in Fig.~\ref{eq:GroupDelay}(b) with that in (d)). The observed increase in group delay with waveguide length is consistent with trends reported for other SIP-supporting geometries and numerical simulations. Nevertheless, for the relatively small and moderately large numbers of unit cells considered here ($N\leq 200$), it is difficult to establish the asymptotic scaling of group delay with $N$ or to distinguish conclusively between contributions from the SIP and other regions of flattened dispersion or band-edge degeneracies. In particular, the upper purple and yellow branches in Fig.~\ref{fig:DispTWG2RA} exhibit relatively flat dispersion and multimode interactions at higher frequencies, between 198 and $199\;\mathrm{THz}$, which may explain the moderate group-delay enhancement observed around $198.6\;\mathrm{THz}$. In contrast, as $N$ increases, the group-delay enhancement near the SIP frequency at $f_{\mathrm{S}}=195.755$ THz becomes progressively more pronounced relative to that at higher frequencies. This behavior is consistent with the stronger length dependence expected for SIP-associated resonances compared with other regions of flattened dispersion, although longer waveguides would be required to verify the asymptotic scaling law.

\section{Conclusion}
\label{ch:Conclusion}
We have presented a systematic design methodology for realizing third-order SIPs in periodic silicon photonic structures comprising three coupled single-mode waveguides with lateral gratings. The proposed approach enables precise engineering of the dispersion diagram to obtain an SIP at a targeted frequency within the optical communication band, while accounting for practical fabrication constraints, including trapezoidal waveguide cross-sections and dimensional tolerances. A key feature of the methodology is the ability to increase the spectral separation between the SIP and neighboring RBEs, thereby facilitating the identification of SIP-associated spectral features in finite-length structures. Numerical tolerance analyses further show that, although geometrical perturbations can alter the exact SIP degeneracy, the waveguides can retain regions of strongly flattened dispersion and correspondingly low group velocity.

Using this methodology, we designed, simulated and fabricated the TWG2RA waveguide and investigated devices comprising $N=50$, $100$, and $200$ unit cells. Comparisons between measured and simulated transmission spectra reveal qualitative agreement, with frequency shifts and discrepancies comparable to those observed in other devices fabricated on the same wafer \cite{furman2025experimentalSIP}. These results provide experimental support for the feasibility of implementing the proposed dispersion-engineered structures using a standard silicon photonics fabrication process. Furthermore, numerical simulations predict an increasingly pronounced group-delay enhancement near the SIP frequency as the number of unit cells increases. For the $N=200$ structure, the simulated group delay reaches approximately $18$ ps, corresponding to an enhancement of about $32$ times relative to a conventional silicon-on-insulator waveguide of the same length. Although direct experimental verification of the SIP and its asymptotic group-delay scaling remains a subject for future investigation, the combined design, fabrication, and transmission results establish a practical route toward integrated slow-light devices exploiting the frozen mode regime at optical communication frequencies.

\section*{Acknowledgment}
This material is based upon work supported by the Air Force Office of Scientific Research (AFOSR) award numbers LRIR 24RYCOR008 and FA9550-18-1-0355, and partial support from the NSF-AFRL Internship program under award ECCS-2030029. The authors additionally thank Dr. R. Marosi, UC Irvine, for contributing to time domain simulations. The authors are thankful to DS SIMULIA for providing CST Studio Suite.

\appendix
\section*{Simulation Mesh Variations}
we highlight the resilience of SIP-related features in the dispersion diagram and finite-length simulations to variations in mesh quality, focusing on the TWG2RA design. While the dispersion diagrams presented previously were obtained using an eigenmode solver with a high-density tetrahedral mesh, here we examine how reducing the mesh density affects the modal behavior. Although variations in mesh resolution do not directly represent fabrication disorder, they introduce numerical perturbations that can provide a preliminary indication of the sensitivity of SIP-related features to small deviations from the ideal model.

\begin{figure}[t]
    \centering
    \begin{subfigure}[t]{0.48\textwidth}
        \centering
        \includegraphics[width=\textwidth]{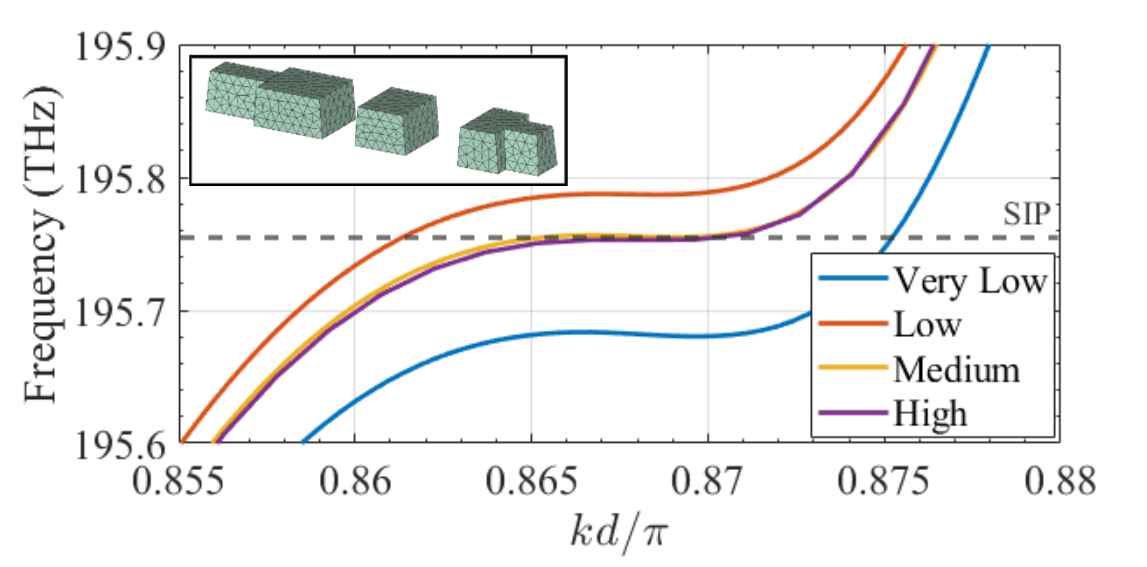}
        \caption{}
    \end{subfigure}\hfill
    \begin{subfigure}[t]{0.48\textwidth}
        \centering
        \includegraphics[width=\textwidth]{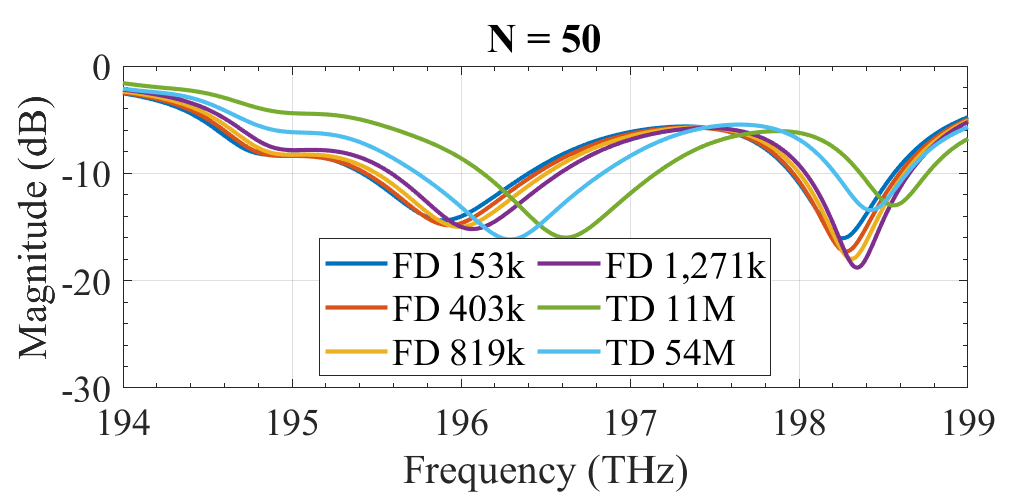}
        \caption{}
    \end{subfigure}
    \caption{(a) Dispersion diagram of the modes in the TWG2RA waveguide near the SIP, obtained using eigenmode simulations with different mesh resolutions. The SIP frequency is indicated by a horizontal line. The inset shows the low-resolution mesh comprising approximately $13{,}000$ mesh cells. (b) Simulated transmission magnitude of the TWG2RA waveguide with $N=50$ unit cells, obtained using frequency-domain (FD) and time-domain (TD) solvers with different mesh resolutions.}
    \label{fig:MeshDifferences}
\end{figure}

In Fig.~\ref{fig:MeshDifferences}(a)  we show the dispersion diagram of the SIP-supporting mode obtained with different mesh resolutions. The Very Low, Low, Medium, and High quality curves correspond to 2k, 13k, 47k, and over 100k mesh cells, respectively.  The inset shows the mesh corresponding to the Low'' setting (13k cells).   As the mesh quality increases, the dispersion curves converge and exhibit the characteristic flattening expected near the SIP. Remarkably, even with relatively coarse meshes, the dispersion curves retain SIP-like features. In particular, both the "Very Low" and "Low" mesh settings produce dispersion curves with SIP-like behavior, although the corresponding frequencies are shifted relative to those obtained with the "Medium" and "High" settings, which are in close agreement. The persistence of relatively flat dispersion, even at low mesh resolutions, indicates that the predicted slow-light behavior can be identified without requiring the highest mesh quality. These results suggest that simulations with coarse meshes can provide useful preliminary estimates of the dispersion characteristics before performing more computationally demanding simulations with finer meshes.

We extend the mesh-quality analysis to finite-length simulations in Fig.~\ref{fig:MeshDifferences}(b). Here, we consider a structure with $N=50$ unit cells and compare four frequency-domain (FD) simulations with two time-domain (TD) simulations performed using CST Studio Suite. As the TD mesh resolution increases, the results approach those obtained with the FD solver. Consistent with the eigenmode simulation results, the four FD simulations exhibit very good agreement despite substantial differences in mesh resolution. In particular, even the relatively coarse FD mesh with 153k cells (dark blue curve) produces results that closely agree with those obtained using the much finer mesh with 1.27 million cells (purple curve), demonstrating the numerical consistency of the simulated transmission response across a wide range of mesh resolutions.

Furthermore, as the mesh quality decreases, the numerical model represents the intended physical geometry with reduced accuracy. This introduces geometrical approximations that are, to some extent, analogous to the differences between the nominal design and fabricated devices. During fabrication, several geometrical parameters may deviate from their intended values, including waveguide width, height, gap size, sidewall angle, and surface roughness. Similarly, a low-quality mesh may introduce inaccuracies in the representation of these geometrical features, resulting in differences between the high-resolution and low-resolution simulations. The persistence of SIP-like dispersion characteristics and transmission features even with relatively coarse meshes therefore provides an encouraging indication of the resilience of the proposed design to geometrical variations. Although mesh-induced numerical errors are not equivalent to fabrication-induced disorder, these results suggest that the fabricated waveguides may retain SIP-like properties despite moderate deviations from the nominal geometry.

These simulations not only suggest a degree of resilience to perturbations but also provide further confidence in the accuracy of the numerical results presented in this work. By varying the mesh quality, we assess the convergence of the simulations and observe very good agreement as the mesh resolution increases. As discussed previously, some differences between fabricated devices and numerical models have been observed in other spectrally sensitive SIP-supporting structures, such as ring resonators and the three-path waveguides reported in \cite{furman2025experimentalSIP}, fabricated on the same wafer. Accordingly, the measured transmission spectra are shifted in frequency to facilitate comparison with the simulated results. Nevertheless, the convergence demonstrated in this appendix supports the numerical reliability of the predicted SIP-related features, while their persistence at lower mesh resolutions provides an additional indication of the design's resilience to certain perturbations.

\ifCLASSOPTIONcaptionsoff
  \newpage
\fi

\bibliographystyle{IEEEtran}
\bibliography{main}

\end{document}